\documentclass[journal]{IEEEtran}
\ifCLASSINFOpdf
\else
\fi
\usepackage{bbm}
\usepackage{verbatim}
\usepackage{graphicx}
\usepackage{cite}
\usepackage{url}
\usepackage[cmex10]{amsmath}
\usepackage{amssymb}
\usepackage{amsmath}
\usepackage{algorithm, algorithmicx, algpseudocode}
\usepackage[caption=false,font=footnotesize]{subfig}
\usepackage{color}
\usepackage{cite}
\usepackage{epstopdf}
\usepackage{calc}
\usepackage{array}
\usepackage{multirow}
\usepackage[printonlyused]{acronym}
\usepackage{cleveref}
\usepackage{siunitx}
\usepackage{mathtools}

\DeclareSIUnit{\belmilliwatt}{Bm}
\DeclareSIUnit{\dBm}{\deci\belmilliwatt}
\DeclareSIUnit{\isotropic}{Bi}
\DeclareSIUnit{\dBi}{\deci\isotropic}
\acrodef{mmW}{millimeter-wave}
\acrodef{BS}{base station}
\acrodef{UE}{user equipment}
\acrodef{SOTA}{state of the art}
\acrodef{AoA}{angle of arrival}
\acrodef{AoD}{angle of departure}
\acrodef{AWV}{antenna weight vector}
\acrodef{ADC}{analog-to-digital converter}
\acrodef{BB}{baseband}
\acrodef{RSRP}{reference signal received power}
\acrodef{CSI}{channel state information}
\acrodef{COTS}{commercial-off-the-shelf}
\acrodef{PAA}{phased antenna array}
\acrodef{TTD}{true-time-delay}
\acrodef{LoS}{line-of-sight}
\acrodef{NLoS}{non-line-of-sight}
\acrodef{IA}{initial access}
\acrodef{DFT}{discrete Fourier transform}
\acrodef{UDN}{ultra-dense networks}
\acrodef{RF}{radio frequency}
\acrodef{MPC}{multipath component}
\acrodef{BF}{beamforming}
\acrodef{SNR}{signal-to-noise ratio}
\acrodef{SINR}{signal-to-interference-plus-noise ratio}
\acrodef{OFDM}{orthogonal frequency-division multiplexing}
\acrodef{DSP}{digital signal processing}
\acrodef{LUT}{lookup table}
\acrodef{MIMO}{multiple-input multiple-output}
\acrodef{IC}{integrated circuits}
\acrodef{PS}{phase shifter}
\acrodef{DAC}{digital-to-analog converter}
\acrodef{EVM}{error vector magnitude}
\acrodef{CP}{cyclic prefix}
\acrodef{FPGA}{field programmable gate arrays}

\newtheorem{proposition}{Proposition}

\DeclareMathOperator*{\argmax}{arg\,max} 

\newcommand{\va}[1]{\mathbf{a}_{#1}}

\newcommand{\tx}[0]{\text{T}}
\newcommand{\rx}[0]{\text{R}}

\newcommand{\hermitian}[0]{\text{H}}
\newcommand{\transpose}[0]{\text{T}}

\newcommand{\Ts}[0]{T_{\text{s}}}

\newcommand{\Mtot}[0]{M}
\newcommand{\Ncip}[0]{N_{\text{cip}}}
\newcommand{\Ncp}[0]{N_{\text{cp}}}

\newcommand{\BW}[0]{\text{BW}}

\newcommand{\taul}[0]{\tau_l}
\newcommand{\taun}[0]{\tilde{\tau}_n}
\newcommand{\fc}[0]{f_{\text{c}}}
\newcommand{\lambdac}[0]{\lambda_{\text{c}}}

\begin{document}
%
\title{Wideband Millimeter-Wave Beam Training with True-Time-Delay Array Architecture}

\author{\IEEEauthorblockN{Han Yan, Veljko Boljanovic, and Danijela Cabric}\\
\IEEEauthorblockA{Department of Electrical and Computer Engineering, University of California, Los Angeles\\
Email: yhaddint@ucla.edu, vboljanovic@ucla.edu, danijela@ee.ucla.edu}
}


\IEEEspecialpapernotice{(Invited Paper)
\vspace{-8mm}
}

\maketitle


\begin{abstract}
Millimeter-wave communications rely on beamforming gain from both transmitters and receivers to compensate for severe propagation loss. To achieve adequate gain, beam training is required to identify propagation directions. The main challenge in beam training arises from maintaining low overhead with increased array size. This paper presents a novel one-shot beam training technique that utilizes the emerging architecture of true-time-delay (TTD) arrays. We first show that TTD arrays facilitate frequency dependent beam steering. The proposed training procedure with TTD arrays then exploits this fact by using a single radio-frequency-chain to multiplex different subcarriers into different sounding directions. We derive conditions on the parameters of TTD array configuration and physical layer to achieve scanning of the entire angular domain with a single orthogonal frequency-division multiplexing (OFDM) training symbol. The estimation of propagation directions with high resolution is achieved via low-complexity digital signal processing of spatially coded subcarriers. Simulation results show that this TTD array based approach requires an order-of-magnitude fewer training symbols than those of phased arrays. 
\end{abstract}


\section{Introduction}
\label{sec:Introduction}

The \ac{mmW} communication is a promising technology for the future cellular network including 5G New Radio (5G-NR). Due to abundant spectrum, it is expected that the mmW network will support ultra-fast data rates. As shown in both theory and prototypes, \ac{mmW} systems requires \ac{BF} with large antenna arrays at both \ac{BS} and \ac{UE} to combat severe propagation loss. The directional beam requires angular channel information for steering direction of analog beams. Such information is typically acquired by beam training in standardized \ac{mmW} systems, including IEEE 802.11ad/ay \cite{8088544} and 5G-NR \cite{dahlman20185g}. However, with increased array size and reduced beam width, the training overhead increases. The challenge of overhead becomes more severe in the future \ac{mmW} and sub-terahertz systems where the array size is expected to further increase due to carrier frequency increase \cite{6G_intro_NYU}.


The \ac{mmW} beam training is an active research area. In the exhaustive sounding scheme, transceivers use one pair of pencil beams at a time. This approach has prohibitive overhead with increased array size. In the multi-stage sounding based scheme, transceivers adapt their beamwidth and pointing angle based on the training results of the previous stage \cite{7845674}. However, its performance degrades with larger path number or user number. In the pseudorandom sounding based scheme, transceivers use pseudorandom beams and compressive sensing to exploit the sparse scattering nature of \ac{mmW} propagation. Although promising results are reported, this approach faces implementation challenges \cite{7400949}.
The above schemes use a single analog beam at a time. Steering multiple beams simultaneously can accelerate beam training. Existing approaches either require to use multiple radio-frequency-chains (RF-chains) \cite{huawei_multibeam}, fully digital architecture \cite{NYU_IA_digital_array}, or utilize leaky wave antenna \cite{leaky_wave_THz}, which introduce additional hardware complexity. Meanwhile, the \ac{TTD} array architecture is an emerging array design for wireless communication systems, particularly in \ac{mmW} and sub-terahertz bands \cite{THz_subarray}. Recent research of \ac{TTD} arrays focus on mitigating squint distortion in beam steering of ultra-wideband signals \cite{USC_UWB_TTD,7394105}. The frequency-controlled beam steering capability of this architecture \cite{1011918} is overlooked and the beam training tailored for \ac{TTD} array is rarely investigated.

In this work, we present the theory and signal processing algorithm of an one-shot beam training scheme using \ac{TTD} array. Firstly, we derive the \ac{TTD} array based \ac{mmW} wideband system model with \ac{CP} based \ac{OFDM} waveforms. We show that \ac{TTD} arrays can apply unique \ac{AWV} to different subcarriers with a single RF-chain. 
Secondly, we analyze the \ac{TTD} array based beam steering and derive the relationship between subcarrier frequencies and the corresponding steering directions in the closed form. The required delay tap spacing of TTD arrays and number of subcarriers to simultaneously scan the entire angular space are also analytically determined. 
Thirdly, we design a novel low-complexity signal processing algorithm to exploit the \ac{TTD} array for beam training. We conduct numerical evaluation using mmMAGIC channel model at 28GHz for both \ac{LoS} and \ac{NLoS} \cite{mmMAGIC_model}. The results show that the proposed approach can achieve accurate beam training using a single OFDM training symbol. 



The rest of the paper is organized as follows. In Section \ref{sec:system_model}, we present the \ac{TTD} array architecture and the mathematical system model for wideband beam training. Section~\ref{sec:codebook} includes the analysis and design of \ac{TTD} array sounding beams for simultaneous angle scan, followed by beam training algorithm in Section~\ref{sec:beam_training_algorithm}. The numerical results are presented in Section~\ref{sec:simulation results}. Finally, Section~\ref{sec:Conclusion} concludes the paper and highlights open research questions.


Scalars, vectors, and matrices are denoted by non-bold, bold lower-case, and bold upper-case letters, respectively.
The $(i,j)$-th element of $\mathbf{A}$ is denoted by $[\mathbf{A}]_{i,j}$. Conjugate, transpose, Hermitian transpose, and pseudoinverse are denoted by $(.)^{*}$, $(.)^{\transpose}$, $(.)^{\hermitian}$, and $(.)^{\dagger}$ respectively. 

%
%
\section{System Model}
\label{sec:system_model}

This section introduces the system model of wideband beam training and problem formulation. 

%
%
Consider a 5G-like single cell system with a \ac{BS} equipped with $N_{\tx}$ antennas. The system utilizes CP-OFDM with sampling duration $\Ts$ (bandwidth $\BW = 1/\Ts$, $\Mtot$ subcarriers, and cyclic prefix length $\Ncp$). The beam training pilots are denoted as $X[m]$, where $m$ is the index of subcarrier. The beam training occurs in the downlink where \ac{BS} transmits beamformed signal over mmW sparse multipath channel to UEs. We focus on the design where UE is equipped with TTD array with single RF-chain and $N_{\rx}$ antennas. We assume a single stream of the training signal transmitted by the BS, that can  use any array architecture\footnote{Typically BS is equipped with multiple RF-chains. We assume they are used for other purposes, e.g., serving other users.}. During sounding, the BS transmitter use \ac{AWV} $\mathbf{v}\in\mathbb{C}^{N_{\tx}}$.


%
%

\begin{figure}
\begin{center}
\includegraphics[width=0.49\textwidth]{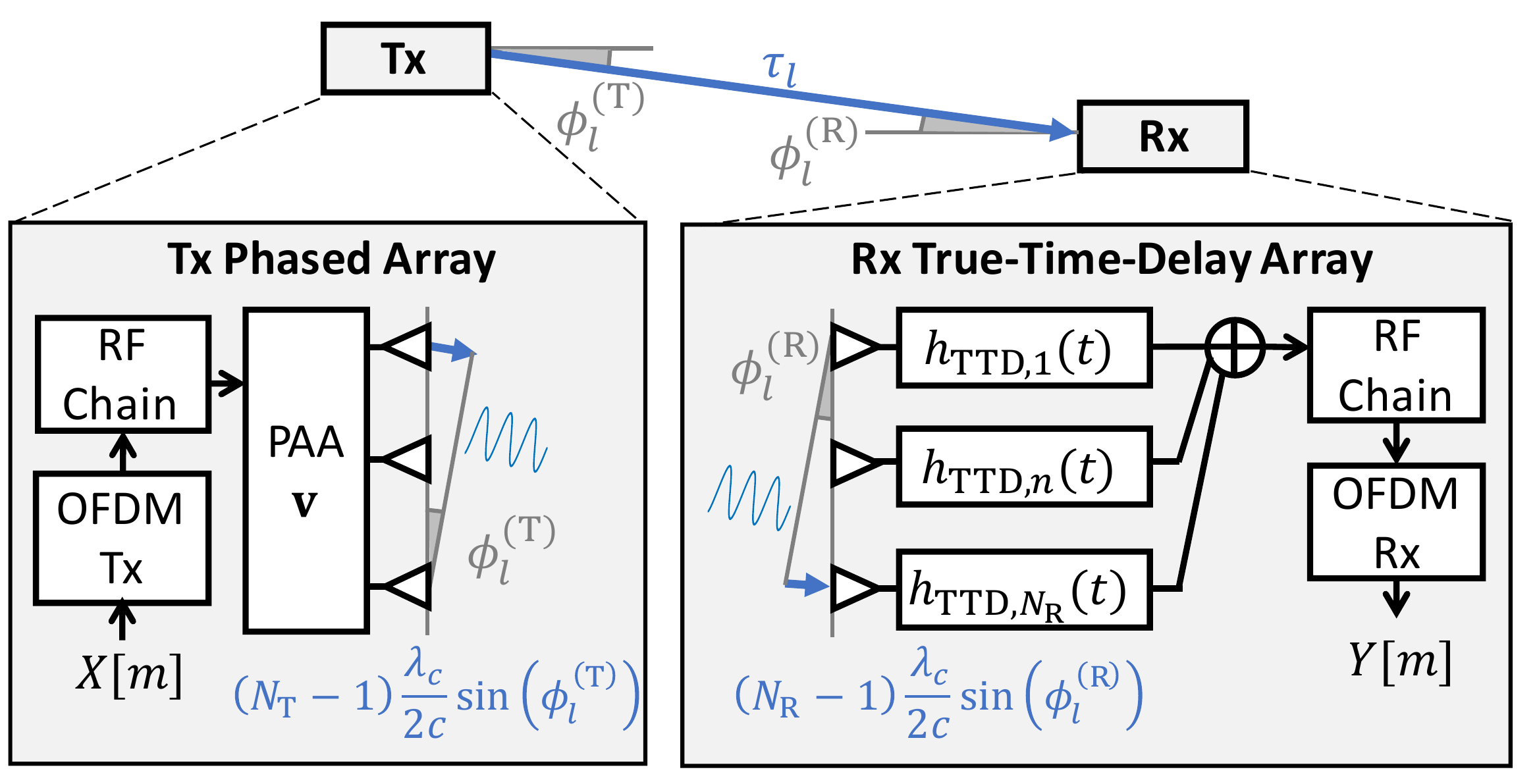}
\end{center}
\vspace{-5mm}
\caption{Illustration of the transceiver and channel model.}
\vspace{-3mm}
\label{fig:system_mode}
\end{figure}

We consider the geometric multipath channel. The time domain channel between the $q$-th transmitter antenna and the $n$-th receiver antenna is denoted as
\begin{align}
\begin{split}
    &h_{q,n}(t)= \rho\sum_{l=1}^{L}g_l p_c\left(t-\Gamma_{l,q,n}\right)
    \end{split}
\end{align}
where $\rho=\sqrt{N_{\tx}N_{\rx}/L}$ is the normalization factor, $L$ is the number of \ac{MPC}\footnote{In the realistic \ac{mmW} channel model, e.g., mmMAGIC \cite{mmMAGIC_model}, the number of \ac{MPC} $L$ is in the order of dozens to hundreds. Although some existing works exploit the clustering nature of \ac{MPC} and deliberately reduce $L$, this work does not require assumption of sparsity.}. Function $p_{\text{c}}(t)$ is the time domain response filter due to limited hardware bandwidth. Scalars $g_{l}$ and $\Gamma_{l,q,n}$ are the complex gain of the $l$-th \ac{MPC} and propagation delay from the $q$-th transmitter element to the $n$-th receiver element via the $l$-th MPC, respectively. In this work, we focus on system where antenna array is linearly arranged with half wavelength spacing (corresponding to center frequency $\fc$). Thus, the delay $\Gamma_{l,q,n}$ is modeled as   
\begin{align*}
    \Gamma_{l,q,n} = \tau_{l} -  \frac{(q-1)\lambda_c\sin\left(\phi_l^{(\tx)}\right)}{2c} + \frac{(n-1)\lambda_c\sin\left(\phi_l^{(\rx)}\right)}{2c}.
\end{align*}
As illustrated in \Cref{fig:system_mode}, $\taul$ is delay between the first transmitter and the first receiver array element due to the $l$-th \ac{MPC}. $\phi^{(\tx)}_{l}$, $\phi^{(\rx)}_{l}$, $\lambdac = c/\fc$ and $c$ are the \ac{AoD} and \ac{AoA} of the $l$-th \ac{MPC}, wavelength of carrier, and the speed of light, respectively.
At the UE receiver, each element introduces time domain thermal noise which is modeled as additive white Gaussian noise $n(t)$ with zero mean and spectral density $N_0/2$.

In principle, the TTD circuit block introduces constant group delay\footnote{Delays can be introduced by transmission line \cite{1011918}. But a more promising approach is to use digitally controlled analog circuits to apply delay either in RF \cite{TTD_RF} or baseband \cite{WSU_TTD}.} to the received signal. Denoting this delay as $\tau_{\text{TTD},n}$ in the $n$-th array element, such module has impulse response 
\begin{align}
    h_{\text{TTD},n}(t) = \delta(t-\tau_{\text{TTD},n}).
\end{align}
%
%
Based on the above model, the received OFDM symbol after analog \ac{TTD} array is determined by the following proposition.

\begin{proposition}
\label{propisition:OFDM_in_TTD}
The received OFDM symbol of the $m$-th subcarrier is
\begin{align}
\begin{split}
Y[m] = \mathbf{w}^{\hermitian}_{\text{TTD}}[m]\mathbf{H}[m]\mathbf{v}X[m] + N[m],
\end{split}
\label{eq:OFDM_received_sig_model}
\end{align}
The combiner specified by \ac{TTD} arrays $\mathbf{w}_{\text{TTD}}[m]\in \mathbb{C}^{N_{\rx}}$ is frequency dependent, i.e., its $n$-th element as
\begin{align}
    [\mathbf{w}_{\text{TTD}}[m]]_n 
    = \mathrm{exp}\left(j2\pi f_m\tau_{\text{TTD},n}\right).
    \label{eq:TTD_AWV}
\end{align}
where the RF frequency of the $m$-th subcarrier is denoted as $f_m$, i.e., 
\begin{align}
    f_m = 
    \begin{cases}
        \fc + \frac{m}{\Mtot}\BW, &0 \leq m< \frac{\Mtot}{2}\\
        \fc -\frac{\BW}{2} + \frac{\left(m-\frac{\Mtot}{2}\right)}{\Mtot}\BW, &\frac{\Mtot}{2} \leq m <\Mtot
    \end{cases}.
\end{align}

The channel at the $m$-th subcarrier $\mathbf{H}[m] \in \mathbb{C}^{N_{\rx} \times N_{\tx}}$ is
\begin{align}
\begin{split}
    \mathbf{H}[m] = \rho\Bigg[&\sum_{l=1}^{L} \tilde{g}_{l}\left(\sum_{i=0}^{\Mtot-1}e^{-j\frac{2\pi im}{\Mtot}} p_{\text{c}}(i\Ts-\tau_{l})\right)\\
    &\cdot\va{\rx}\left(\theta^{(\rx)}_{l},f_m\right) \va{\tx}^{\hermitian}\left(\theta^{(\tx)}_{l},f_m\right)\Bigg].
    \end{split}
    \label{eq:channel_frequency_domain}
\end{align}
In the above equation, $\tilde{g}_l = g_l\mathrm{exp}(-j2\pi \fc \tau_{l})$ is the channel gain. The array responses $\mathbf{a}_{\rx}(\phi^{(\rx)}_l,f_m)$ and $\mathbf{a}_{\tx}(\phi^{(\tx)}_l,f_m)$ are defined by their $n$-th and $q$-th element as
\begin{align*}
\left[\mathbf{a}_{\rx}\left(\phi^{(\rx)}_l,f_m\right)\right]_n 
= &\mathrm{exp}\left[-j\pi (n-1)(f_m /\fc)\sin\left(\phi^{(\rx)}_l\right)\right],\\
\left[\mathbf{a}_{\tx}\left(\phi_l^{(\tx)},f_m\right)\right]_q 
= &\mathrm{exp}\left[-j\pi (q-1)(f_m /\fc)\sin\left(\phi^{(\tx)}_l\right)\right].
\end{align*} 

This model holds true so long as the \ac{CP} is longer than the cumulative delay of both \ac{MPC} and \ac{TTD} circuits, i.e.,
\begin{align}
\Ncp \Ts > \max_{l,q,n} \Gamma_{l,q,n} + \max_n \tau_{\text{TTD},n},
\label{eq:CP_condition}
\end{align}

The frequency domain noise $N[m]$ is Gaussian distributed with zero mean and variance $\mathbb{E}|N[m]|^2 = N_0\BW/(2\Mtot)$. 
\end{proposition}

\begin{IEEEproof}
See Appendix. \ref{appendix:OFDM_in_TTD}
\end{IEEEproof}
\textit{Remark1:} Following the derivation in Appendix\ref{appendix:OFDM_in_TTD}, it is straightforward to generalize \textit{Proposition 1} to system where transmitter also uses TTD array. 

In this work, we focus on the system design where TTD array introduces uniformly spaced delay taps during beam training, i.e., $\tau_{\text{TTD},n} = (n-1)\Delta\tau$ where $\Delta\tau$ denotes the delay spacing and it meets condition $\Delta\tau > 1/(2\fc)$. 
%
%
%
With straightforward mathematical manipulation, the receiver beamforming gain at direction $\alpha$ for the subcarrier with frequency $f_m$ can be expressed $ G(\alpha,f_m) = N_{\rx}^{-1}|\mathbf{w}_{\text{TTD}}^{\hermitian}[m]\mathbf{a}_{\rx}(\alpha,f_m)|^2$ 
\begin{align}
    G(\theta,f_m) = 
    \frac{1}{N_{\rx}} \left|\frac{\sin\left[\frac{N_{\rx}\pi}{2}\left(2f_m\Delta\tau+\left(f_m/\fc\right)\sin(\theta)\right)\right]}{\sin\left[\frac{\pi}{2}\left(2f_m\Delta\tau+\left(f_m/\fc\right)\sin(\theta)\right)\right]}\right|^2.
    \label{eq:G_function}
\end{align}

In this work, we assume that pilots subcarrier have non-zero power when subcarrier indices are in set $\mathcal{M}$, i.e., $X[m]\neq 0, m \in \mathcal{M}$. Further, the transmit pilots have unit power constraint as
$\sum_{m=0}^{\Mtot-1}|X[m]|^2 = \Mtot$.
We address two problems in TTD array based beam training.

\textit{Problem 1 (\ac{TTD} sounding beams design):} The objective is to design delay tap spacing $\Delta\tau$ and number of subcarriers $\Mtot$ such that the sounding beams of subcarriers scan the entire angular region. In other words, for arbitrary \ac{AoA} $\theta$, there is at least one subcarrier which has sufficient beamforming gain in its direction. Mathematically, our goal is to find the feasible set $\mathcal{S}$ of design parameters such that sounding beamforming gain does not drop below $(1-\epsilon)N_{\rx}$ for the least favorable \ac{AoA} $\theta$:
\begin{align}
    \mathcal{S} = \left\{\left(\Delta\tau, \Mtot\right) \big| \min_{\theta}\max_{m}G(\theta,f_m) \geq (1-\epsilon)N_{\rx}\right\}.
\end{align}


\textit{Problem 2 (TTD based beam training):} 
Using the design parameters from $\mathcal{S}$, the objective is to design beam training pilots $\mathcal{M}$ and signal processing algorithm to estimate dominant propagation directions from a single received symbol $Y[m]$. In this work, we focus on the receiver beam training and assume transmitter beamformer $\mathbf{v}$ has been aligned with \ac{AoD}.
%
%
\section{TTD based beam training design}
In this section, we present the TTD based sounding beam design and beam training algorithm.

\subsection{Analysis and design of TTD sounding beams}
\label{sec:codebook}

We first analyze the frequency dependent receiver gain function (\ref{eq:G_function}). Denote $\Psi= 2f_m\Delta\tau+\left(f_m/\fc\right)\sin(\theta)$, the following observations are made \cite[Chapt 7.2.4]{tse2005fundamentals}.
\begin{itemize}
    \item $G(\Psi)$ has a peak at $\Psi = 0$.
    \item $G(\Psi)$ is a periodic function of $\Psi$ with period 2.
\end{itemize}
Based on the above properties, the center of the sounding beam corresponding to subcarrier $f_m$, i.e.,
$\theta_m^{\star} = \argmax_{\theta} G(\theta,f_m)$, is given in the following proposition.
\begin{proposition}
Given delay tap spacing $\Delta\tau$, an approximation of the pointing direction of sounding beam that corresponds to the $m$-th subcarrier is
\begin{align}
\theta^{\star}_m \approx \sin^{-1}\left(\mathrm{mod}(2f_m \Delta\tau+ 1, 2) - 1 \right),
\label{eq:angle_mapping_approximate}
\end{align}
where $\mathrm{mod}()$ is the modulo operation.
\end{proposition}

\begin{IEEEproof}
This is by solving $\Psi = 2z,z\in\mathbb{Z}$ with given $\Delta\tau$ and $f_m$. The closed form solution (\ref{eq:angle_mapping_approximate}) is available by using the approximation $f_m/\fc \approx 1,\forall m$ in $\Psi$.
\end{IEEEproof}

Next, we discuss the design of delay tap spacing $\Delta\tau$ and number of subcarriers $\Mtot$ for a given $\BW$ in \textit{problem 1}. For this purpose, we define $\epsilon$-beamwidth as $\Omega(\epsilon,N_{\rx})$ such that \begin{align}
    G(\Psi) \geq (1-\epsilon)N_{\rx}, \quad \Psi \in [-\Omega(\epsilon,N_{\rx}),  \Omega(\epsilon,N_{\rx})]
\end{align}
for a given required sounding gain factor $1-\epsilon$ and receiver array size $N_{\rx}$. 
Note that the specific value of  $\Omega(\epsilon,N_{\rx})$ can be found numerically, e.g., 3dB-beamwidth is $\Omega(0.5,N_{\rx}) = 0.886/N_{\rx}$ \cite[Chapt 22.7]{array_textbook1}. Note that directly solving \textit{problem 1} is challenging. In the following proposition, we show a subset of the solution of \textit{problem 1}.
\begin{proposition}
A subset of $\mathcal{S}$, denoted as $\mathcal{S}_s$, is 
\begin{align}
    \mathcal{S}_s = \left\{\left(\Delta\tau,\Mtot\right)\bigg| \Delta\tau \geq \frac{1}{\BW}  + \frac{1}{2\fc}, \Mtot \geq \frac{\BW\Delta\tau+\frac{\BW}{2\fc}}{\Omega(\epsilon,N_{\rx})} \right\}.
    \label{eq:critical_dtau}
\end{align}
\end{proposition}

\begin{IEEEproof}
See Appendix~\ref{appendix:condition_of_S}.
\end{IEEEproof}

In the system where $\BW\ll \fc$, the condition on delay tap spacing $\Delta\tau$ can be relaxed and the critical value is $\Delta\tau = 1/\BW$. With such delay tap spacing, the critical value of required number of subcarriers is $M=\lceil1/\Omega(\epsilon,N_{\rx})\rceil$ where ceiling operator $\lceil a \rceil$ gives the smallest integer that is greater than $a$. 

\subsection{TTD based beam training}
\label{sec:beam_training_algorithm}

Using the proposed TTD sounding beams, the beam training is greatly simplified. Effectively, UE receiver only needs to configure TTD array based on predefined delay tap spacing $\Delta\tau$ during the scheduled time slot for beam training. A \ac{LUT} can be constructed based on (\ref{eq:angle_mapping_approximate}), which contains the pointing direction of all training pilots $m\in\mathcal{M}$ and sounding directions. During training, the UE measures the \ac{RSRP} of the pilots, i.e., 
\begin{align}
    m_{\text{best}} = \argmax_m  |Y[m]|^2
\end{align} and use the index of subcarrier that has highest \ac{RSRP} $m_{\text{best}}$ and \ac{LUT} (\ref{eq:angle_mapping_approximate}) to estimate \ac{AoA}, i.e.,
\begin{align}
    \hat{\phi}^{(\rx)} = \theta^{\star}_{m_{\text{best}}}.
\end{align}

The proposed TTD array beam training approach has an interesting relationship with \ac{PAA} receiver beam training, when TTD delay tap spacing $\Delta\tau = 1/\BW$ and $\fc\Delta\tau \in \mathbb{Z}$. In fact, the proposed simultaneous multi-beam can exactly represent \ac{DFT} and oversampled \ac{DFT} sounding beams, which is commonly used by \ac{PAA} receiver. In the \ac{DFT} beam based training procedure, the \ac{PAA} receiver uses \ac{AWV} 
\begin{align}
    [\mathbf{w}_{\text{PAA},k}]_n = \mathrm{exp}[j2\pi (n-1)k/K]
    \label{eq:PAA_sounding}
\end{align}
for the $k$-th training symbol, and it requires a total $K$ \ac{OFDM} training symbols (usually $K$ is greater than $N_{\rx}$ and it is power of 2). Comparing (\ref{eq:TTD_AWV}) and (\ref{eq:PAA_sounding}), it is straightforward that the TTD receiver can mimic this procedure by utilizing $\Mtot=K$ subcarriers in a single \ac{OFDM} training symbol.

It is worth noting that the total number of subcarriers $\Mtot$ is in the order of magnitude of receiver array size $N_{\rx}$ when system intends to obey minimum condition in \textit{proposition 3} or mimic \ac{PAA} beam training. However, the total number of subcarrier in practical \ac{mmW} system needs also accommodate coherent bandwidth, and therefore it can be much larger than $N_{\rx}$. As such, the TTD array based system needs to use \textit{subcarrier selection} scheme. Based on (\ref{eq:TTD_AWV}), the following two \ac{TTD} systems are identical: 1) system has a total number of subcarrier $M$ and training utilizes all subcarriers $m< \Mtot$; 2) system has a total number of subcarrier $\Mtot R$ for some $R\in\mathbb{Z}$ and training utilizes subcarriers within set $\mathcal{M} = \{mR|0 \leq m< \Mtot\}$. With this feature, the proposed system is directly extendable to existing \ac{mmW} protocols.


    

%
%
\section{Numerical Results}
\label{sec:simulation results}
This section presents the numerical results of the beam pattern and the performance of the proposed beam training. 

\begin{figure}
\begin{center}
\includegraphics[width=0.5\textwidth]{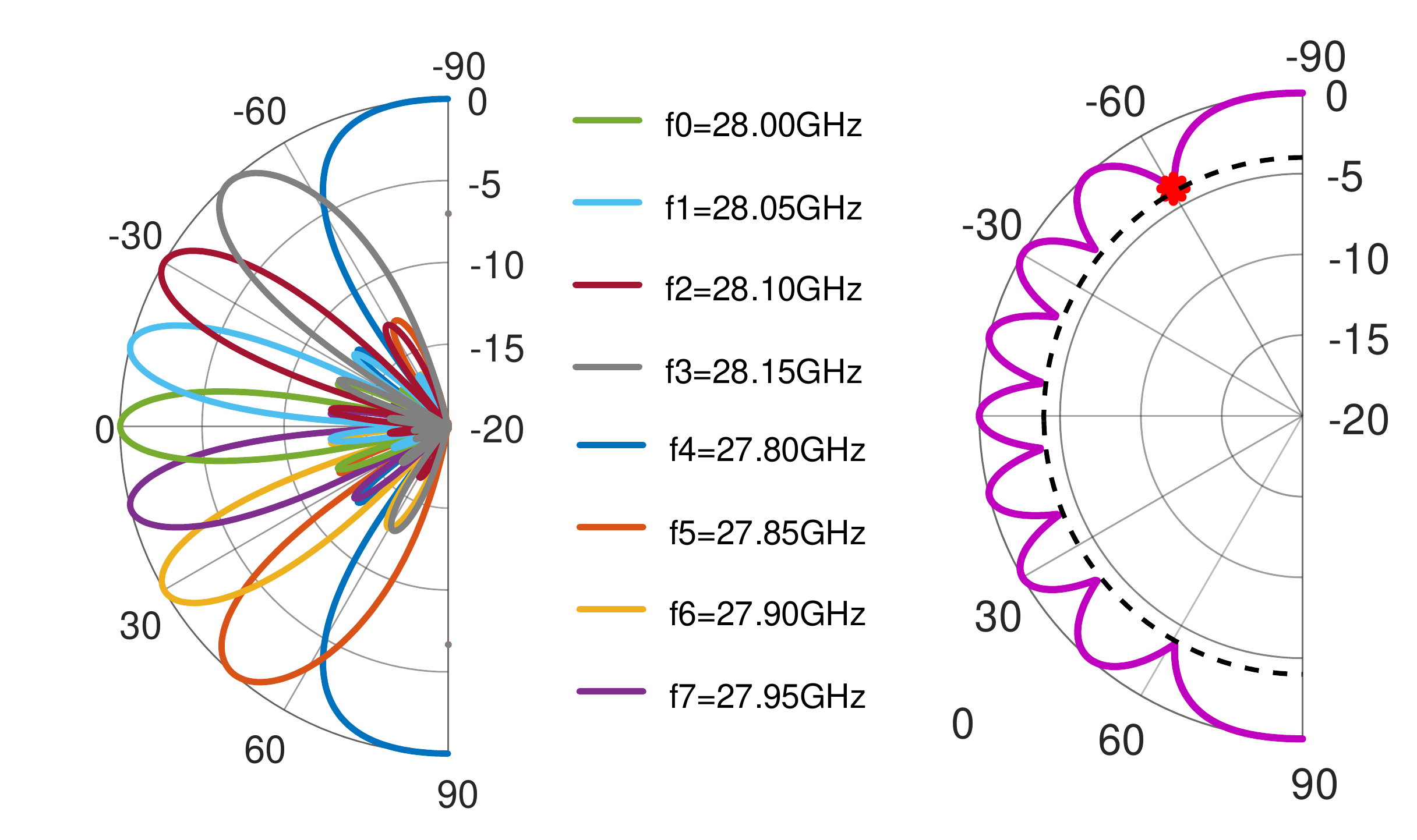}
\end{center}
\vspace{-5mm}
\caption{Example of TTD array based frequency dependent beam pattern in the logarithmic scale. In the left figure, different colored curves show beam patterns $G(\theta,f_m)$ for subcarriers $f_m$. In the right figure, the purple curve, the red circle, and the black dashed curve illustrate the function $\max_{m}G(\theta,f_m)$, $\min_{\theta}\max_{m}G(\theta,f_m)$, and gain constraint $1-\epsilon$, respectively.}
\vspace{-2mm}
\label{fig:TTD_beam_pattern}
\end{figure}
%
%
\Cref{fig:TTD_beam_pattern} shows an example of the TTD array based simultaneous multi-beam and design parameters that follow \textit{Proposition 3}. The simulation utilizes carrier frequency $\fc=$\SI{28}{\giga\hertz}, bandwidth \SI{400}{\mega\hertz}, delay tap spacing $\Delta\tau=$\SI{2.5}{\nano\second}, and $N_\rx = 8$ receiver antennas. The number of subcarrier $\Mtot=8$ is designed based on constraint $1-\epsilon=0.4$ (\SI{4}{\dB} loss), the corresponding $\delta(\epsilon,N_\rx) = 0.1266$, and $\Mtot = \lceil1/\delta(\epsilon,N)\rceil$. The figure shows that the gain of sounding beams meet the constraint for the least favorable \ac{AoA}. It also verifies the relationship between subcarrier frequencies and the center of sounding beams in (\ref{eq:angle_mapping_approximate}).


%
%

\begin{figure}
\begin{center}
\includegraphics[width=0.5\textwidth]{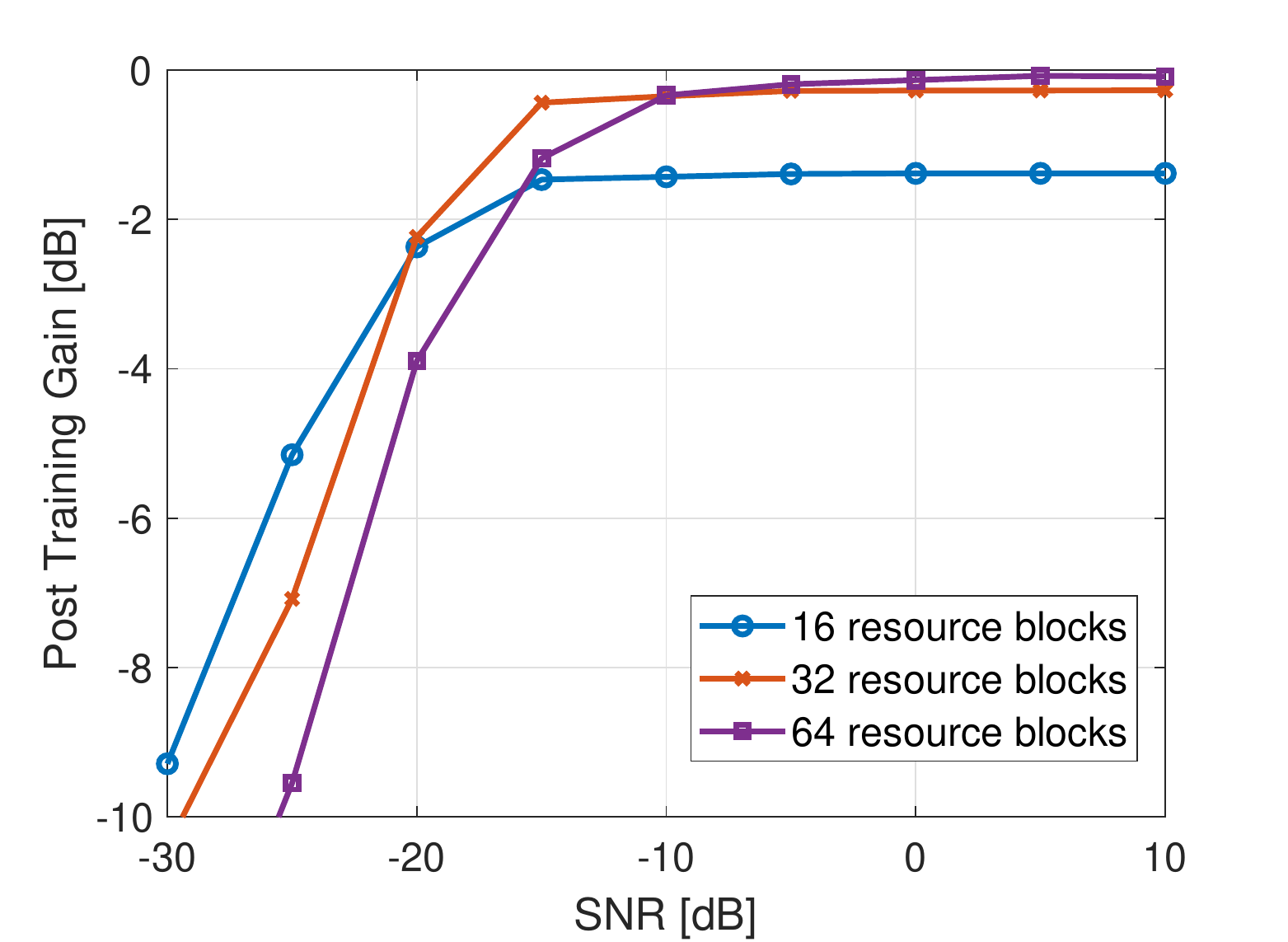}
\end{center}
\vspace{-4mm}
\caption{Post-beam-training gain of the proposed TTD beam training algorithm in pure \ac{LoS} environment.}
\vspace{-4mm}
\label{fig:single_ray_channel}
\end{figure}

The beam training performance of the proposed BF training algorithm in a pure \ac{LoS} channel, i.e., number of MPC is $L=1$, is presented in \Cref{fig:single_ray_channel}. The simulation parameters are identical to previous case, except the system has $N_{\tx}=64$ transmitter antennas and $N_\rx = 16$ receiver antennas. Furthermore, a total of $\Mtot=2048$ subcarriers are used. The \ac{SNR} is defined as post-transmitter-beam \ac{SNR} across all band, i.e., 
\begin{align*}
\mathrm{SNR} = \frac{\sum_{m}\|\mathbf{H}[m]\mathbf{v}\|^2}{\sum_{m}\mathbb{E}|N[m]|^2}.
\end{align*}
The simulation follows the concept of frequency \textit{resource block} in cellular systems \cite{dahlman20185g}. Namely, when a subcarrier is used as the pilot, the 12 neighbor subcarriers are also used. The beam training utilizes only a single \ac{OFDM} symbol. We use normalized post training beamforming gain as a metric, i.e., $N_{\rx}^{-2}|\mathbf{a}^{\hermitian}_{\rx}(\hat{\phi}^{(\rx)})\mathbf{a}_{\rx}({\phi}_1^{(\rx)})|^2$. The figure shows that a smaller number of used subcarriers provides improvement in the low SNR regime, because more power is loaded in the corresponding pilots. However, it limits the training resolution of the proposed algorithm, and therefore the post training gain saturates in high SNR regime. Using an increased number of subcarriers improves performance in the high SNR regime, but the benefit saturates when the sounding beam number is more than $2N_{\rx}$.

\begin{figure}
\begin{center}
\vspace{-4mm}
\includegraphics[width=0.48\textwidth]{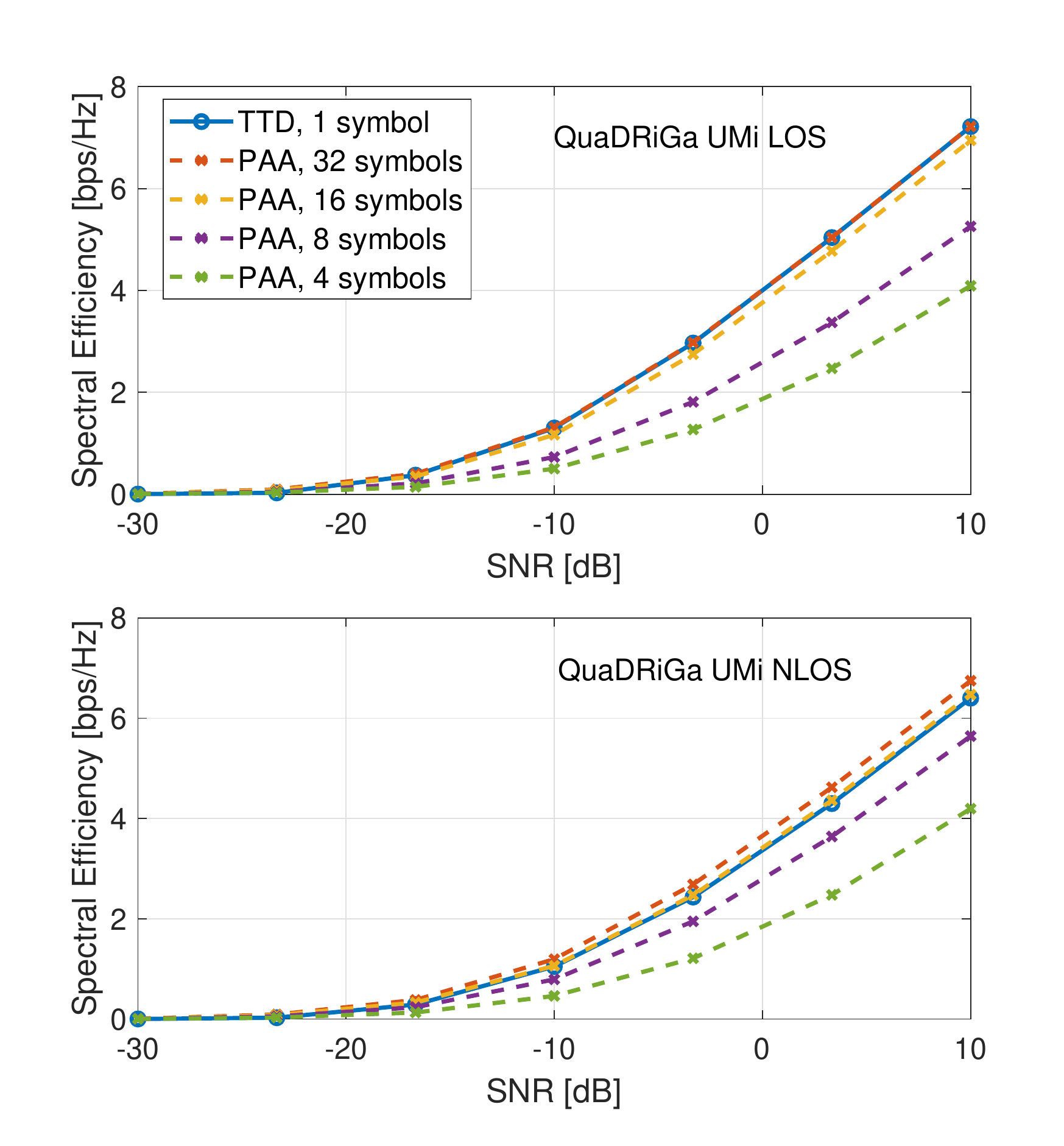}
\end{center}
\vspace{-8mm}
\caption{Post-training spectral efficiency comparison between the proposed TTD system and the \ac{PAA} \ac{DFT} sounding system.}
\vspace{-4mm}
\label{fig:quadriga_sim}
\end{figure}

The post-beam-training spectral efficiency is shown in \Cref{fig:quadriga_sim}. 
Particularly, the evaluation utilizes QuaDRiGa simulator \cite{6758357} with mmMAGIC 28GHz channel model \cite{mmMAGIC_model} in urban micro (UMi) \ac{LoS} and \ac{NLoS} environments. The number of MPC $L$ in these two scenarios are 41 and 79, respectively. In the TTD array based system, 32 frequency \textit{resource blocks} and 1 OFDM symbol are used for beam training. A benchmark system that uses a \ac{PAA} receiver with same array geometry is used for comparison. The \ac{PAA} receiver utilizes beam training \ac{AWV} from sampled columns of a $32$ by $32$ \ac{DFT} matrix when receiving the training symbols. The same frequency resources and power are used in the training pilots for both systems. In the \ac{LoS} environment, the proposed system is equivalent to the benchmark using $2N_{\rx}$ training symbols. Severe performance degradation is observed in the \ac{PAA} with small number of training symbols, since the \ac{DFT} beams fail to scan entire angular domain. In the \ac{NLoS} environment, the proposed TTD array beam training method is outperformed by the \ac{PAA} with the same angular coverage, i.e., using $32$ symbols, by a small margin. This is because \ac{PAA} provides robustness to the frequency selective channel gain when conducting wideband power measurement during beam training. 
In summary, the evaluation in realistic \ac{mmW} propagation channels reveals that the proposed method can use order of magnitude smaller number of training symbols as compared to \ac{PAA} with \ac{DFT} sounding beams.

%
%
\section{Conclusions and Future works}
\label{sec:Conclusion}
In this work, we present a fast beam training scheme that utilizes true-time-delay based millimeter-wave array. Exploiting the frequency dependence of this array architecture's antenna weight vector, the system can simultaneously scan multiple angles via frequency resources, and thus greatly accelerate the beam training procedure. We derive the condition for delay tap spacing and the required number of subcarriers. Based on these novel sounding beams, a one-shot, low-complex beam training algorithm is developed. The simulation results reveal that the proposed method utilizes a single training symbol to complete beam training. Such feature is appealing in the future \ac{mmW} systems where conventional phased array based beam training meet increased training overhead due to increased array size.

There are many open questions in \ac{TTD} array based signal processing and transceiver design. Firstly, the feasibility of \ac{TTD} array based simultaneous multi-beam in 3D environment with planar array remains unknown. Secondly, it is of interest to exploit the frequency dependent \ac{AWV} of \ac{TTD} arrays to develop algorithms for super-resolution beam training, channel estimation, and covariance estimation in \ac{mmW} wideband system. Lastly, it is of critical importance to understand the impact and required specification of hardware in RF and mixed signal domain for power efficient operation of \ac{TTD} arrays. 

\section{Acknowledgement}
This work was supported in part by NSF under grant 1718742. This work was also supported in part by the ComSenTer and CONIX Research Centers, two of six centers in JUMP, a Semiconductor Research Corporation (SRC) program sponsored by DARPA. The authors would like to thank Dr. Subhanshu Gupta, Dr. Deuk Heo, and Erfan Ghaderi from Washington State University for helpful discussions.

\appendix
\subsection{Proof of proposition 1}
\label{appendix:OFDM_in_TTD}

\begin{figure*}[ht]
\begin{align}
\begin{split}
&[\mathcal{F}(\check{\mathbf{h}})]_m
=\sum_{l=1}^{L}\sum_{n=1}^{N_{\rx}}\sum_{q=1}^{N_{\tx}}\sum_{i=0}^{\Mtot -1}e^{-j\frac{2\pi mi}{\Mtot}}\left[g_{l}p\left(i\Ts-\taul - \frac{(n-1)\lambdac\sin\left(\phi^{(\rx)}_l\right)}{c} + \frac{(q-1)\lambdac\sin\left(\phi^{(\tx)}_l\right)}{c} 
-\taun\right)e^{j\theta_{l,q,n}}\right]\\
=&\sum_{l=1}^{L}\sum_{n=1}^{N_{\rx}}\sum_{q=1}^{N_{\tx}}\left[\sum_{i=0}^{\Mtot -1}e^{-j\frac{2\pi mi}{\Mtot}}g_{l}p\left(i\Ts-\taul\right)\right]e^{-j\frac{2\pi m (n-1)\lambdac\sin\left(\phi_l\right)}{\Ts \Mtot c}}e^{j\frac{2\pi m (q-1)\lambdac\sin\left(\phi_l\right)}{\Ts \Mtot c}}e^{-j\frac{2\pi m\taun}{\Ts\Mtot}}e^{j\theta_{l,q,n}}\\
=&\sum_{l=1}^{L}\Bigg\{\underbrace{g_{l}e^{-j2\pi \fc \taul}}_{\tilde{g}_l}\left[\sum_{i=0}^{\Mtot -1}e^{-j\frac{2\pi mi}{\Mtot}}p\left(i\Ts-\taul\right)\right]\sum_{n=1}^{N_{\rx}}\underbrace{\left(e^{-j\frac{2\pi m\taun}{\Ts\Mtot}}e^{-j2\pi \fc \taun}\right)}_{\left[\mathbf{w}^{*}[m]\right]_n}\underbrace{\left(e^{-j\frac{2\pi m (n-1)\lambdac\sin(\phi_l)}{\Ts c\Mtot}}e^{-j\frac{2\pi \fc (n-1)\lambdac\sin(\phi_l)}{2c}}\right)}_{[\mathbf{a}_{\rx}(\phi_l,f_m)]_n}\\
& \cdot\sum_{q=1}^{N_{\tx}}\underbrace{\left(e^{j\frac{2\pi m (q-1)\lambdac\sin(\phi_l)}{\Ts c\Mtot}}e^{j\frac{2\pi \fc (q-1)\lambdac\sin(\phi_l)}{2c}}\right)}_{[\mathbf{a}^{*}_{\tx}(\phi_l,f_m)]_q}\underbrace{e^{j\theta_q}}_{[\mathbf{v}]_q}\Bigg\}
\end{split}
\label{eq:OFDM_proof_big}
\end{align}
\hrulefill
\end{figure*}

Consider the continuous time channel between the $q$-th transmitter and $n$-th receiver element due to the $l$-th MPC \textit{before} and \textit{after} the TTD circuits as $\tilde{h}^{(\text{RF})}_{l,q,n}(t)$ and $h^{(\text{RF})}_{l,q,n}(t)$, respectively. These two systems are modeled as 
$\tilde{h}^{(\text{RF})}_{l,q,n}(t) = g_l p_c\left(t-\Gamma_{l,q,n}\right)$
and
$h^{(\text{RF})}_{l,q,n}(t) = \tilde{h}_{l,q,n}(t) * h_{\text{TTD},n}(t) = g_lp_c(t-\tau_{l,q,n})$,
where $*$ is the convolution operator. For notation clarity, we denote $\tau_{l,q,n} = \Gamma_{l,q,n} + \tilde{\tau}_n$ and $\tilde{\tau}_n \triangleq \tau_{\text{TTD},n}$.

For a real pass-band signal $x(t) = \Re[s(t)e^{j2\pi ft}]$, the complex baseband equivalent model of the impulse response is written as
\begin{align}
\begin{split}
    h_{l,q,n}(t) = g_lp_c(t-\tau_{l,q,n})e^{j\theta_{l,q,n}},
\end{split}
\end{align}
where $\theta_{l,q,n} \triangleq -2\pi \fc \tau_{l,q,n} + \theta_q$ and $\theta_q = \angle[\mathbf{v}]_q$ is the phase shift value introduced by the $q$-th transmitter circuit. 

The discrete time channel with sampling duration $\Ts$ is denoted as 
\begin{align}
\begin{split}
    h_{l,q,n}[i] = &h_{l,q,n}(i\Ts)
    = g_lp_c(iT_s-\tau_{l,q,n})e^{j\theta_{l,q,n}}.
\end{split}
\end{align}
The overall channel response is then written in vector form by $\mathbf{h}_{l,n} \in \mathbb{C}^{\Ncip}$ where
$[\mathbf{h}_{l,n}]_i = h_{l,n}[i]$. Note that $\Ncip\Ts \geq \max_{l,q,n}\Gamma_{l,q,n} + \max_n \tau_{\text{TTD},n}$.

The time domain OFDM sample sequence after sampling of combined $N_\rx$ antenna signals can be written as matrix form 
\begin{align}
    \mathbf{y} = \sum_{n=1}^{N_{\rx}}\sum_{l=1}^{L}\mathbf{H}_{l,n}\tilde{\mathbf{x}}
\end{align}
where $\mathbf{y} = [y[\Mtot-1], y[\Mtot-2], \cdots, y[0]]^{\transpose} \in \mathbb{C}^{\Mtot}$ is the received signal in the time domain that corresponds to one OFDM symbol. $\tilde{\mathbf{x}} = [x[\Mtot-1], \cdots, x[0], x[-1], \cdots, x[-\Ncp]]^{\transpose} \in \mathbb{C}^{\Mtot+\Ncp}$ is the transmits time domain signal after both CP and signal. Channel $\mathbf{H}_{l,n} \in \mathbb{C}^{\Mtot \times (\Mtot + \Ncp)}$ is a cyclic matrix for $1 \leq n \leq N_{\rx}$ according to the linear convolution of channel \cite[Chapt 12.22]{goldsmith2005wireless}. 
The specific expression of this channel matrix for the $n$-th antenna element ($1 \leq n \leq N_\rx$) by its $k$-th row ($1 \leq k \leq \Mtot$) as
\begin{align}
    [\mathbf{H}_{l,n}]_{k,:}=[
\mathbf{0}^{\transpose}_{k-1}, \mathbf{h}^{\transpose}_{l,n}, \mathbf{0}^{\transpose}_{\Mtot+\Ncp - \Ncip-k+1}].
\end{align}
Note that we generalized the definition of all-zero vector $\mathbf{0}_n\in \mathbb{C}^{n}$ such that $\mathbf{0}_0$ is empty. Further, for valid dimension to exist in above equation, the CP length needs to follow condition $\Ncp \geq \Ncip$, i.e., (\ref{eq:CP_condition}).

Next, we discuss the the effective channel after CP removal. Due to the fact that both TTD analog combining and CP removal are linear operations, the cyclic matrix that corresponds digital baseband after CP-removal is denoted as 
$\check{\mathbf{H}} \in \mathbb{C}^{\Mtot \times \Mtot}$.
This cyclic matrix is defined by its first row $\check{\mathbf{h}}^{\transpose}$ as
\begin{align}
\begin{split}
\check{\mathbf{h}}^{\transpose} = [\check{\mathbf{H}}]_{1,:} = \left[\sum_{n=1}^{N_{\rx}}\sum_{l=1}^{L}\mathbf{h}_{l,n}, \mathbf{0}_{\Mtot -\Ncip}^{\transpose}\right]
\end{split}
\end{align}


Note that the frequency domain channel that takes account analog precoder and combiner is the eigenvalues of $\check{\mathbf{H}}$. Due to the property of cyclic matrix, it can be achieved by taking discrete Fourier transform $\mathcal{F}()$ of the row vector $\check{\mathbf{h}}^{\transpose}$. Particularly, its $m$-th component is denoted as in (\ref{eq:OFDM_proof_big}), where the second equality sign is due to the \textit{sampling theorem} and relationship between \textit{discrete time Fourier transform} and \textit{discrete Fourier transform}. Therefore (\ref{eq:OFDM_proof_big}) shows the post-beamformer channel is $\mathbf{w}^{\hermitian}_{\text{TTD}}[m]\mathbf{H}[m]\mathbf{v}$ as in (\ref{eq:OFDM_received_sig_model}).

Further, the TTD operation $h_{\text{TTD},n}(t)$ does not change power spectral density of thermal noise in each element, and therefore the noise term is $N[m]$ defined in (\ref{eq:OFDM_received_sig_model}).


\subsection{Proof of proposition 3}
\label{appendix:condition_of_S}
Let define set $\mathcal{P} = \{\Psi|G(\Psi) > (1-\epsilon)N_{\rx}\}$. From properties of function (\ref{eq:G_function}), this set is equivalent to $\mathcal{P} = \{\Psi|\min_{z\in\mathbb{Z}}|\Psi - 2z| \leq \delta(\epsilon,N_{\rx})\}$. Thus, the set in terms of $f$ and $\theta$ such that $\mathcal{P} = \{(f_m,\theta)|G(\theta,f_m)>N_\rx(1-\epsilon)\}$ can be written as
\begin{align*}
    \mathcal{P} = \left\{\left(f_m,\theta\right)\bigg|\min_{z\in \mathbb{Z}} \left|2z + 2f_m\Delta\tau + \frac{f_m \sin(\theta)}{\fc}\right| \leq \delta(\epsilon,N_{\rx}) \right\}.
\end{align*}
Therefore, \textit{problem 1} becomes to find set $\mathcal{S}$ such that
\begin{align*}
    \mathcal{S} =
    \left\{\left(\Delta\tau,\Mtot\right)\bigg|\max_\theta \min_{z\in\mathbb{Z},m < \Mtot} \left|z +\frac{m \xi}{\Mtot}  -  C_\theta\right| \leq \frac{\delta(\epsilon,N_{\rx})}{2} \right\},
\end{align*}
where we define parameter $C_\theta = -\fc\Delta\tau - \sin(\theta)/2$ and $\xi = \BW[\Delta\tau + \sin(\theta)/(2\fc)]$ for clarity.

Due to the degree of freedom of integer $z$, the condition function in the previous set is equivalent to 
\begin{align}
    \max_{\theta}\min_{m< \Mtot} \left|\frac{m}{\Mtot} \xi - \tilde{C}_\theta\right| \leq \frac{\delta(\epsilon,N_{\rx})}{2}
    \label{eq:condition_in_proof}
\end{align}
where $\tilde{C}_\theta \in [0,1)$. Denote $f_{\text{quan}}(m)= m\xi/\Mtot$ as a quantifier with range $\xi$ and granularity $\xi/\Mtot$. A sufficient condition for (\ref{eq:condition_in_proof}) to be valid is that $f_{\text{quan}}(m)$ has its range greater than 1 and its granularity less than $\delta(\epsilon,N_{\rx})$, i.e.,
\begin{align}
    \xi \geq 1, \forall \theta \quad \text{and} \quad \xi/\Mtot \leq \delta(\epsilon,N_{\rx}), \forall \theta 
    \label{eq:condition_in_appendix}
\end{align}
which leads to $\mathcal{S}_s$ of (\ref{eq:critical_dtau}).


\bibliographystyle{IEEEtran}
\bibliography{IEEEabrv,references}

\end{document}